%% file: main.tex
\documentclass[sigconf]{acmart}

\usepackage{booktabs} 
\usepackage{subfigure}
\usepackage{bm}
\usepackage{xspace}
\usepackage{color}
\usepackage{multirow}
\usepackage{footmisc}
\usepackage{pifont}
\newcommand{\paratitle}[1]{\vspace{1.5ex}\noindent\textbf{#1}}
\newcommand{\ie}{\emph{i.e.,}\xspace}

\newcommand{\eg}{\emph{e.g.,}\xspace}

\newcommand{\ignore}[1]{}

\copyrightyear{2020} 
\acmYear{2020} 
\setcopyright{acmcopyright}\acmConference[CIKM '20]{Proceedings of the 29th ACM International Conference on Information and Knowledge Management}{October 19--23, 2020}{Virtual Event, Ireland}
\acmBooktitle{Proceedings of the 29th ACM International Conference on Information and Knowledge Management (CIKM '20), October 19--23, 2020, Virtual Event, Ireland}
\acmPrice{15.00}
\acmDOI{10.1145/3340531.3412095}
\acmISBN{978-1-4503-6859-9/20/10}

\ignore{
\copyrightyear{2020} 
\acmYear{2020} 
\setcopyright{acmcopyright}\acmConference[CIKM '20]{Proceedings of the 29th ACM International Conference on Information and Knowledge Management}{October 19--23, 2020}{Virtual Event, Ireland}
\acmBooktitle{Proceedings of the 29th ACM International Conference on Information and Knowledge Management (CIKM '20), October 19--23, 2020, Virtual Event, Ireland}
\acmPrice{15.00}
\acmDOI{10.1145/3340531.3412095}
\acmISBN{978-1-4503-6859-9/20/10}
}

\begin{document}
\title{Revisiting Alternative Experimental Settings for Evaluating Top-$N$ Item Recommendation Algorithms}

\author{Wayne Xin Zhao$^{1,2}$, Junhua Chen$^{3}$, Pengfei Wang$^{4*}$, Qi Gu$^{3}$ and Ji-Rong Wen$^{1,2}$}\thanks{$^*$Corresponding author.}
\affiliation{%
 \institution{$^1$Gaoling School of Artificial Intelligence, Renmin University of China}
 \institution{$^2$Beijing Key Laboratory of Big Data Management and Analysis Methods}
 \institution{$^3$School of Information, Renmin University of China}
 \institution{$^4$ Beijing University of Posts and Telecommunications}
}
\affiliation{%
  \institution{ \{batmanfly, cjh1507, guqi@ruc.edu.cn, jrwen\}@ruc.edu.cn,~wangpengfei@bupt.edu.cn}
}

\ignore{
\author[1]{Wayne Xin Zhao}
\additionalaffiliation{%
\department[1]{School of Information} 
 \institution{Renmin University of China}
}
\affiliation{%
\department[0]{Gaoling School of Artificial Intelligence} 
 \institution{ Renmin University of China}
}
\email{	batmanfly@gmail.com}
\author{Junhua Chen}
\affiliation{%
\department[1]{School of Information} 
 \institution{Renmin University of China}
}
\email{chovychen@tencent.com}
\author{Pengfei Wang}
	\authornote{Corresponding author.}
\affiliation{%
  \institution{Beijing University of Posts and Telecommunications}
}
\email{wangpengfei@bupt.edu.cn}
\author{Qi Gu}
\affiliation{%
\department[0]{School of Information} 
 \institution{ Renmin University of China}
}
\email{guqi@ruc.edu.cn}
\author[rvt]{Ji-Rong Wen}
\additionalaffiliation{%
\department[1]{School of Information} 
 \institution{Renmin University of China}
}
\affiliation{%
\department[0]{Gaoling School of Artificial Intelligence} 
 \institution{ Renmin University of China}
}
\email{	jrwen@ruc.edu.cn}
}

\begin{abstract}
Top-$N$ item recommendation has been a widely studied task from implicit feedback. Although 
much progress has been made with neural methods, there is increasing concern on appropriate  evaluation  
of  recommendation algorithms.  
In this paper, we revisit alternative experimental settings for evaluating top-$N$ recommendation algorithms, considering three important factors, namely 
\emph{dataset splitting}, \emph{sampled metrics} and \emph{domain selection}.
We select eight representative recommendation algorithms (covering both traditional and neural methods) 
and construct extensive  experiments on a very large dataset.
By carefully revisiting different options, we make several important findings on the three factors, which directly provide useful suggestions on how to appropriately  set up the experiments for top-$N$ item recommendation.

\end{abstract}


\maketitle


\input{sec-intro}

\input{sec-exp}

\section{Conclusion}
In this paper, we empirically compared different experimental settings of three important factors for   top-$N$ item recommendation.  Our  experiments have led to  several empirical suggestions for evaluating  item recommendation algorithms.
First, for dataset splitting, random ordering with ratio-based splitting is the suggested option for evaluating time-insensitive algorithms, while leave-one-out splitting can be applied to small datasets. 
Second, we should be careful to use sampled metrics. If it was used, we  suggest  sampling a large number of items. Third, it is suggested to use multiple  datasets  from diverse domains as evaluation sets. 
As future work,  we will consider constructing online evaluation  for studying the effect of various factors. Also, more factors and datasets will be investigated for evaluating recommendation algorithms. 

\section{Acknowledgements}
This research work was partially supported by the National Natural Science Foundation of China under Grant No. 61872369, 61832017 and 61802029,  Beijing Academy of Artificial Intelligence~(BAAI) under Grant No.BAAI2020ZJ0301, and Beijing Outstanding Young Scientist Program under Grant No. BJJWZYJH012019100020098.

\bibliographystyle{elsarticle-num-names}
\bibliography{ref}

\end{document}

%% file: sec-intro.tex
\section{Introduction}
In the past decade, top-$N$ item recommendation has been a widely studied task from implicit feedback~\cite{bpr},  which aims to identify a small set of  items that a user may prefer from a large collection. 
Various top-$N$ recommendation algorithms have been developed, specially the great progress made with deep learning~\cite{survey}. 

To prove the effectiveness of a recommendation algorithm, one needs to construct reliable evaluation experiments  on benchmark datasets. 
Typically, such an evaluation procedure consists of a series of setup steps on datasets, metrics, baselines and other protocols. As each setup step can be conducted with different options, it is essential to develop  and design appropriate criterions for standardizing the experimental settings~\cite{comparison,Thiago2019}. 


In the literature,  a number of studies have been proposed to standardize  evaluation criterions for top-$N$ item recommendation~\cite{Thiago2019,Harald2013,overall,comparison}.
However, they mainly adopt traditional recommendation algorithms as evaluation targets.
It is not clear whether some specific finding still holds when neural algorithms are involved in evaluation. 
As another limit,  prior studies  may not well respond to recent concerns~\cite{Rendle2020} about evaluation protocols on neural recommendation algorithms. 
The studied or compared settings in \cite{Thiago2019,Harald2013,overall, comparison}  do not align with the major divergence from current debate. Besides, existing studies usually use very few comparison methods or datasets. 
Therefore, there is a need to thoroughly revisit   experimental settings of substantial divergence in recent literature, considering both traditional and neural methods.





In this paper, we present a large-scale empirical study on the effect of different experimental settings for  
 evaluating top-$N$ item recommendation algorithms. 
 We try to identify  important evaluation settings that have led to major divergence in recent progress~\cite{survey,Rendle2020}.  
In specific, we  consider three important influencing factors, namely dataset splitting, sampled metrics and domain selection.
\emph{Dataset splitting} refers to the strategy to construct training, validation and test sets using original data; \emph{sampled metrics} refers to the strategy to compute the metric results with sampled irrelevant items; and \emph{domain selection} refers to the strategy to select suitable datasets 
from different domains for evaluation.

  To examine the effect of  the three factors, we  construct extensive experiments on the Amazon review dataset~\cite{amazon16}, containing 142.8 million user-item interaction records from 24 domains.
Different from prior works~\cite{comparison,overall},  which analyze how each individual method performs under different settings, we study how one selected factor affects the overall performance ranking of different comparison methods, since top-$N$ item recommendation is essentially a ranking task. 
We select eight representative recommendation algorithms as comparison methods, including both traditional and neural methods.
We utilize three ranking correlation measures to quantitatively characterize such ranking differences.

 
\ignore{To examine the effect of  the three factors, we  construct extensive experiments on the Amazon review dataset~\cite{amazon16}, which contains 142.8 million user-item interaction records from 23  domains. Furthermore, we select eight representative top-$N$ recommendation algorithms as comparison methods, including four traditional and four neural methods. Our core idea is to study the varying of the performance ranking over the eight comparison methods under different configurations (\ie a combination of different settings for the studied factors). We incorporate three ranking correlation measures to quantitatively characterize such ranking differences.
}

Our empirical study has lead to the following findings.
First, for dataset splitting, temporal ordering seems to yield a substantially different performance ranking compared with random ordering. 
An appropriate option should depend on the specific task. 
 A suggestion is to adopt random ordering in a general setting while temporal ordering for time-sensitive cases~(\eg sequential recommendation). 
Interestingly, the simple yet widely adopted  leave-one-out splitting strategy has a significant correlation with  ratio-based splitting strategy.
It can be used for small datasets.  
 Second, the performance ranking based on sampled metrics has a relatively weak correlation with the exact ranking, and increasing the number of sampled items will improve the correlation degree. When using sampled metrics, researchers should use a large number of sampled irrelevant items as possible. 
 Finally,  data domains with varying domain characteristics or sparsity levels are likely to yield substantially different performance rankings. A good strategy is to select representative datasets that are able to cover different aspects of multiple domains.

%% file: sec-exp.tex
\section{Experimental Setup}

In this section, we set up the experiments by describing the datasets, comparison methods, and evaluation metrics. 

\paratitle{Datasets}. We adopt the Amazon product review dataset~\cite{amazon16} for evaluation, containing  142.8 million reviews from 24 domains. 
For 
top-$N$ recommendation, each review is considered as an \emph{interaction record} between a user and an item, while the rest information is discarded, \eg text and metadata. 
Since several comparison methods cannot obtain a result in a reasonable time on the largest  \emph{book} domain, we remove this domain for the efficiency issue.
User-item interaction data from the rest 23 domains as the final dataset. 
We further adopt the released \emph{five-core copies} of the original review dataset to remove inactive users or infrequent items. 

\paratitle{Comparison Methods}. 
We adopt eight recommendation algorithms, including popularity, ItemKNN, SVD++~\cite{Koren2008} and BPR~\cite{bpr}, 
DSSM~\cite{Huang2013}, NCF~\cite{He2017}, DIN~\cite{DIN} and GCMC~\cite{gcmc}.
\ignore{
We first consider eight commonly used recommendation algorithms in research community, namely popularity, ItemKNN, SVD++~\cite{Koren2008} and BPR~\cite{bpr}, 
DSSM~\cite{Huang2013} and NCF~\cite{He2017}. Among these six methods, 
popularity and ItemKNN are mainly  based on simple global or item-specific statistics,  SVD++ and BPR utilize matrix factorization techniques, and DSSM  and NCF characterize user-item interactions by using neural networks. 
Besides, we incorporate two recent neural methods from industry community, namely  DIN~\cite{Zhou2019} and its enhanced version DIEN~\cite{Zhou2019}, which learn  user preference  by attending to historical behaviors.
}
Among these eight methods, 
popularity and ItemKNN are mainly on based simple global or item-specific statistics,  SVD++ and BPR utilize matrix factorization techniques,  DSSM  and NCF characterize user-item interactions by using neural networks, DIN learns  user preference  by attending to existing behaviors, and GCMC adopts graph neural networks for recommendation.
The eight methods have a good coverage of  traditional and neural approaches.
We  adopt either original or official implementations for these methods. 
In this paper, we only consider general item recommendation instead of other tasks such as context-aware and sequential recommendation. 
Note that our focus is not to identify the best  algorithm, but study how different experimental settings affect the final performance rankings. 

\paratitle{Evaluation Metrics}. Top-$N$ item recommendation can be considered as a ranking task, in which the recommendations at top positions are important to consider. Following \cite{Jonathan2004,Thiago2019}, we use four metrics in following experiments: (1)~truncated Precision and Recall at top $K$ positions (P$@K$ and R$@K$), (2)~Mean Average Precision (MAP), and (3)~Area under the ROC curve (AUC).
We also computed the results for another two metrics of nDCG$@K$ and MRR. 
They yield the similar results with the above four metrics and omitted. 

\section{Experimental Protocol}
In this section, we present the  experimental protocol for comparing experimental settings for top-$N$ item recommendation. 

\paratitle{Configuration}. We introduce the term ``\emph{configuration}'' to denote a kind of combination for different options of the three factors, namely \emph{dataset splitting}, \emph{sampled metrics}  and  \emph{domain selection}. 
We select the three factors because there is still substantial divergence (lacking standardized discussion) in recent literature of neural methods.
Note that we would not enumerate all possible options for the three factors. Instead, 
we only consider  popular or controversial options from recent studies~\cite{survey}.
In order to reduce the  influence of other factors, we either report results separately by different options or set them to the suggested option by prior studies~\cite{Thiago2019,Harald2013,overall,comparison}. 




\paratitle{Correlation Measurement}. Given a configuration, we can obtain a ranked list of the eight comparison methods, called \emph{performance ranking},  according to the descending order of their performance based on some metric.
We adopt  three measures to quantitatively characterize the correlation or similarity degree between two performance rankings: (1) \textit{Overlap Ratio at top-$K$ positions}~(OR@$k$) computing the overlap ratio of top $k$ methods between two ranked lists; (2) \textit{Spearman's Rank Correlation}~(SRC)  measuring the association between two ranked lists; and (3) ~\textit{Inversion Pair Count}~(IPC) counting the number of inversion pairs between two ranked lists.
The reason to select the three measures is given as follows. 
First, correlation measures (\eg Spearman or Kendall) are commonly used to reflect ranking differences in prior evaluation studies~\cite{overall}. Second, IPC  provides a more intuitive 
understanding  on the values of SRC. Finally, for  item recommendation, top positions are more important to consider, which is captured by OR@$k$. 

\paratitle{Procedure Overview}. 
Given a metric, we first derive a performance ranking  of the eight methods according to some configuration (optimized with validation sets). 
To examine the effect of one factor, we will accordingly generate multiple configurations by considering the alternative options.
Then, we compute the correlation degree between the performance rankings under two different configurations using the above measures.
Finally, the correlation results will be averaged over 23 data domains (except Section 4.3). 

\begin{figure}[!t]
\centering
\includegraphics[width=0.9\columnwidth]{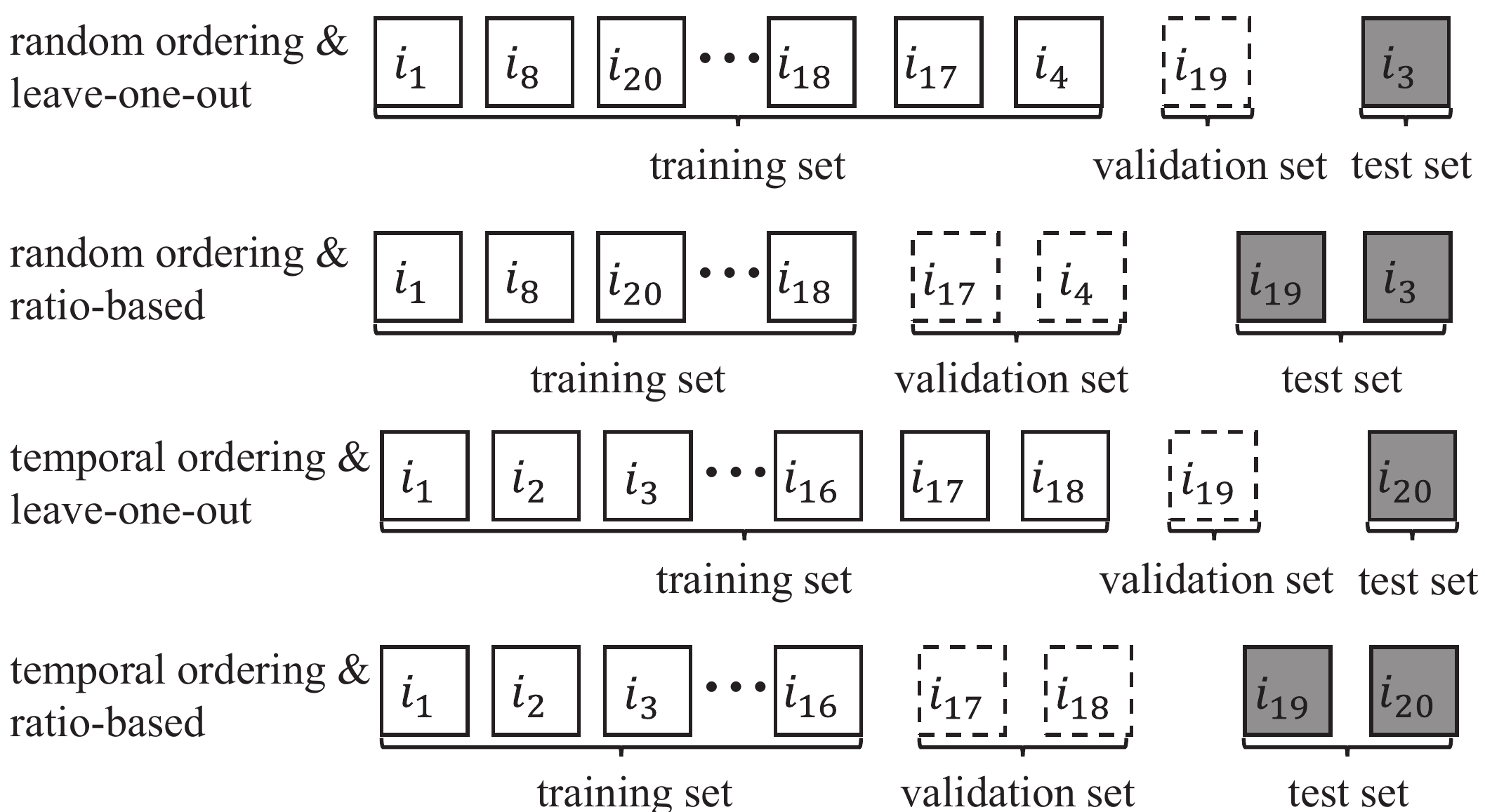}
\caption{An illustrative example for four splitting strategies on a sample user. The user has interacted with
 twenty items.
The subscript of an item $i$  denotes the interaction order with the user: a smaller index indicates an earlier interaction time. We use normal, dash-lined and grey boxes to denote the training, validation and  test sets, respectively. }\label{fig:example}
\end{figure}

\section{Experiment}
In this section, we present the experiment results related to the three factors, namely \emph{dataset splitting}, 
\emph{sampled metrics} and \emph{domain selection}.
When considering one factor, we fix the rest two factors. 
That is to say, given two configurations to compare, we only vary the studied factor, while the rest settings will be set to the same in two compared configurations. 

\subsection{Analysis on Dataset Splitting}
We first study the effect of  different dataset splitting strategies (\ie constructing  training/validation/test sets) on performance ranking. 

\paratitle{Setting}. For each user, we first organize the interaction records of a user using two methods:
(1)~\underline{\textit{Random Ordering~(RO)}} randomly shuffles the items;
(2)~\underline{\textit{Temporal Ordering~(TO)}}  sorts the items according to their interaction timestamps. 
Then, the reordered user-item interaction sequences can be split using two common methods:  
(1)~\underline{\textit{Ratio-based Splitting~(RS)}}  splits the dataset into  three parts for training, validation and test according to a predefined ratio, which  is set to 8:1:1 here. We repeat the  process for five times  to generate different evaluation sets for computing  average results. 
(2)~\underline{\textit{Leave-one-out Splitting~(LS)}}  selects one ground-truth item as test set and another one as validation set, while the rest items are considered as training set.
\emph{LS} strategy can be considered as a special case for \emph{RS}, where both validation and test sets contain only one item. 
Considering both ordering and splitting, we generate four combinations in total, which are illustrated in Figure~\ref{fig:example}. 

\paratitle{Results}.  We present the  comparison results between two different  configurations in Table~\ref{t:data_segmentation_result}. 
First, compared with the splitting method (either \emph{ratio} or \emph{leave-one-out}), the item ordering  way (either \emph{random} or \emph{temporal}) seems to have a more significant influence on the performance ranking. For each metric, the correlation values from the first two lines are substantially weaker  than the last two lines. With  temporal ordering, it is essentially  similar to the setting of sequential recommendation. 
The results indicate that the option of item ordering should depend on specific recommendation tasks. 
It is suggested to adopt \emph{random ordering} in a general setting (especially for evaluating time-insensitive recommendation algorithms), while adopt  \emph{temporal ordering} in time-sensitive cases (\eg sequential recommendation).
Second, with the same item ordering way, the two splitting methods yield very similar ranked lists (see the last two lines for each metric). 
Indeed, leave-one-out splitting has been widely adopted in recent literature~\cite{He2017,amazon16}. 
We suggest using ratio-based splitting when possible for more accurate evaluation.
 While, leave-one-out splitting seems to be preferred with small datasets, since we can use more training data for alleviating  data sparsity.

\begin{table}[!t]
\centering
\caption{Correlation comparison with different configurations on dataset splitting. 
The  results are averaged over the 23 domains (with standard deviations). Here, ``RS'' and ``LS'' denote ratio-based or leave-one-out splitting;  ``RO'' and ``TO'' denote  random or temporal ordering. ``$\uparrow$'' (``$\downarrow$'') indicates a larger (smaller) result is better.
 }
\label{t:data_segmentation_result}
\footnotesize
\setlength{\tabcolsep}{1.5mm}{
\begin{tabular}{ccccc}
\toprule
Metrics& Comparison & OR@3 $(\uparrow)$ & SRC $(\uparrow)$ & IPC $(\downarrow)$ \\
\midrule
\multirow{4}{*}{P@10}&(\textit{$TO,RS$}) \emph{v.s.} (\textit{$RO,RS$}) & $0.731_{\pm0.232}$ & $0.687_{\pm0.221}$ & $3.588_{\pm2.042}$ \\
&(\textit{$TO,LS$}) \emph{v.s.} (\textit{$RO,LS$}) & $0.757_{\pm0.198}$ & $0.709_{\pm0.237}$ & $2.826_{\pm2.306}$ \\
&(\textit{$RO,LS$}) \emph{v.s.} (\textit{$RO,RS$}) & $0.916_{\pm0.144}$ & $0.875_{\pm0.155}$ & $1.212_{\pm0.994}$ \\
&(\textit{$TO,LS$}) \emph{v.s.} (\textit{$TO,RS$}) & $0.815_{\pm0.199}$ & $0.755_{\pm0.184}$ & $1.660_{\pm1.762}$ \\
\midrule
\multirow{4}{*}{R@10}& (\textit{$TO,RS$}) \emph{v.s.} (\textit{$RO,RS$}) & $0.727_{\pm0.266}$ & $0.705_{\pm0.274}$ & $3.499_{\pm2.418}$ \\
&(\textit{$TO,LS$}) \emph{v.s.} (\textit{$RO,LS$}) & $0.698_{\pm0.221}$ & $0.649_{\pm0.243}$ & $3.866_{\pm2.557}$ \\
&(\textit{$RO,LS$}) \emph{v.s.} (\textit{$RO,RS$}) & $0.901_{\pm0.152}$ & $0.872_{\pm0.144}$ & $1.650_{\pm1.464}$ \\
&(\textit{$TO,LS$}) \emph{v.s.} (\textit{$TO,RS$}) & $0.828_{\pm0.190}$ & $0.769_{\pm0.242}$ & $2.039_{\pm1.282}$ \\
\midrule
\multirow{4}{*}{AUC}&(\textit{$TO,RS$}) \emph{v.s.} (\textit{$RO,RS$}) & $0.727_{\pm0.195}$ & $0.702_{\pm0.190}$ & $3.168_{\pm2.281}$ \\
&(\textit{$TO,LS$}) \emph{v.s.} (\textit{$RO,LS$}) & $0.797_{\pm0.244}$ & $0.715_{\pm0.235}$ & $2.644_{\pm2.299}$ \\
&(\textit{$RO,LS$}) \emph{v.s.} (\textit{$RO,RS$}) & $0.915_{\pm0.147}$ & $0.825_{\pm0.277}$ & $1.431_{\pm1.124}$ \\
&(\textit{$TO,LS$}) \emph{v.s.} (\textit{$TO,RS$})& $0.905_{\pm0.151}$ & $0.823_{\pm0.199}$ & $1.170_{\pm1.148}$ \\
\midrule
\multirow{4}{*}{MAP}&(\textit{$TO,RS$}) \emph{v.s.} (\textit{$RO,RS$}) & $0.769_{\pm0.234}$ & $0.762_{\pm0.265}$ & $2.909_{\pm2.337}$ \\
&(\textit{$TO,LS$}) \emph{v.s.} (\textit{$RO,LS$}) & $0.757_{\pm0.239}$ & $0.696_{\pm0.252}$ & $3.212_{\pm2.677}$ \\
&(\textit{$RO,LS$}) \emph{v.s.} (\textit{$RO,RS$}) & $0.930_{\pm0.135}$ & $0.911_{\pm0.134}$ & $1.130_{\pm1.288}$ \\
&(\textit{$TO,LS$}) \emph{v.s.} (\textit{$TO,RS$}) & $0.884_{\pm0.182}$ & $0.799_{\pm0.172}$ & $2.088_{\pm1.900}$ \\
\bottomrule
\end{tabular}}
\end{table}

\subsection{Analysis on Sampled Metrics}

Next, we study the effect of  sampled metrics (\ie only a smaller set of sampled items and the ground-truth items are ranked for computing the metrics) on performance ranking.

\paratitle{Setting}. For test, it is time-consuming to take all the items from the item set as the candidate when the size of item set is large. An alternative way is to  sample a small set of items as irrelevant items. Then, the ground-truth items and sampled items  are merged as a single candidate list for  ranking. The results of the metrics will be computed based  on such an item subset.
This way is called \emph{sampled metrics}~\cite{Rendle2020}.
We consider two sampling strategies, namely \emph{uniform sampling} and \emph{popularity-biased sampling}, which samples irrelevant items according to either a uniform distribution or a frequency-based distribution. We further consider using three different numbers of irrelevant samples, namely $\{10,50,100\}$, which means that a ground-truth item will be paired with 10, 50 or 100 sampled items. 
When we adopt leave-one-out splitting, the case becomes \emph{real-plus-$N$}~\cite{overall,comparison}.  
 For comparison, we  adopt the entire item  set (excluding  ground-truth items) for ranking (denoted by $all$) as a referencing setting.
Following Section 4.1, for dataset splitting, we adopt  ratio-based dataset splitting (denoted by $RS$) with random ordering (denoted by $RO$) in all compared configurations.
 

\begin{table}[!t]
\centering
\footnotesize
\caption{Correlation comparison with different configurations on sampled metrics. The  results are averaged  over  23 domains.   ``$\Delta=\{RO, RS\}$''  denotes  random ordering and  ratio-based splitting method for dataset splitting,  ``pop'' / ``uni'' denote  popularity/uniform sampling,  the  subscript denotes the number of sampled items and ``all'' denotes  all non-ground-truth items.
 }\label{t:analysis_sampling}
\setlength{\tabcolsep}{1.5mm}{
\begin{tabular}{ccccc}
\toprule
 Metric & Comparison & {OR@3} ($\uparrow$) & {SRC} ($\uparrow$) & {IPC} ($\downarrow$)\\
\midrule
\multirow{4}{*}{P@10}
&$(\Delta,all)$ $\emph{v.s.}$ $(\Delta,pop_{10})$ & $0.531_{\pm0.302}$ & $0.566_{\pm0.285}$ & $6.111_{\pm3.971}$ \\
&$(\Delta,all)$ $\emph{v.s.}$ $(\Delta,pop_{50})$ & $0.577_{\pm0.297}$ & $0.604_{\pm0.300}$ & $5.799_{\pm3.289}$\\
&$(\Delta,all)$ $\emph{v.s.}$ $(\Delta,pop_{100})$ & $0.709_{\pm0.207}$ & $0.606_{\pm0.328}$ & $5.223_{\pm2.168}$ \\
&$(\Delta,all)$ $\emph{v.s.}$ $(\Delta,uni_{10})$ & $0.678_{\pm0.349}$ & $0.667_{\pm0.329}$ &
 $3.644_{\pm2.801}$\\
&$(\Delta,all)$ $\emph{v.s.}$ $(\Delta,uni_{50})$ & $0.777_{\pm0.229}$ & $0.722_{\pm0.254}$ & $2.966_{\pm2.424}$\\
&$(\Delta,all)$ $\emph{v.s.}$ $(\Delta,uni_{100})$ & $0.872_{\pm0.189}$ & $0.804_{\pm0.178}$ & $2.357_{\pm1.200}$ \\
\midrule
\multirow{4}{*}{R@10}
&$(\Delta,all)$ $\emph{v.s.}$ $(\Delta,pop_{10})$ & $0.581_{\pm0.305}$ & $0.544_{\pm0.292}$ & $6.670_{\pm2.949}$ \\
&$(\Delta,all)$ $\emph{v.s.}$ $(\Delta,pop_{50})$ & $0.667_{\pm0.231}$ & $0.607_{\pm0.273}$ & $5.711_{\pm2.302}$ \\
&$(\Delta,all)$ $\emph{v.s.}$ $(\Delta,pop_{100})$ & $0.712_{\pm0.200}$ & $0.619_{\pm0.318}$ & $5.143_{\pm2.166}$ \\
&$(\Delta,all)$ $\emph{v.s.}$ $(\Delta,uni_{10})$ & $0.801_{\pm0.161}$ & $0.695_{\pm0.212}$ & $2.579_{\pm1.672}$ \\
&$(\Delta,all)$ $\emph{v.s.}$ $(\Delta,uni_{50})$ & $0.864_{\pm0.194}$ & $0.788_{\pm0.177}$ & $1.891_{\pm1.621}$ \\
&$(\Delta,all)$ $\emph{v.s.}$ $(\Delta,uni_{100})$ & $0.872_{\pm0.189}$ & $0.804_{\pm0.173}$ & $2.164_{\pm1.215}$ \\
\midrule
\multirow{4}{*}{AUC}
&$(\Delta,all)$ $\emph{v.s.}$ $(\Delta,pop_{10})$ & $0.659_{\pm0.272}$ & $0.690_{\pm0.233}$ & $5.379_{\pm2.852}$ \\
&$(\Delta,all)$ $\emph{v.s.}$ $(\Delta,pop_{50})$ & $0.673_{\pm0.281}$ & $0.726_{\pm0.205}$ & $5.066_{\pm2.399}$ \\
&$(\Delta,all)$ $\emph{v.s.}$ $(\Delta,pop_{100})$ & $0.695_{\pm0.278}$ & $0.737_{\pm0.197}$ & $4.922_{\pm2.311}$ \\
&$(\Delta,all)$ $\emph{v.s.}$ $(\Delta,uni_{10})$ & $0.844_{\pm0.224}$ & $0.764_{\pm0.253}$ & $1.994_{\pm2.295}$ \\
&$(\Delta,all)$ $\emph{v.s.}$ $(\Delta,uni_{50})$ & $0.868_{\pm0.210}$ & $0.793_{\pm0.247}$ & $1.893_{\pm2.290}$ \\
&$(\Delta,all)$ $\emph{v.s.}$ $(\Delta,uni_{100})$ & $0.878_{\pm0.181}$ & $0.813_{\pm0.230}$ & $1.629_{\pm1.644}$ \\
\midrule
\multirow{4}{*}{MAP}
&$(\Delta,all)$ $\emph{v.s.}$ $(\Delta,pop_{10})$ & $0.599_{\pm0.315}$ & $0.559_{\pm0.288}$ & $5.780_{\pm2.159}$ \\
&$(\Delta,all)$ $\emph{v.s.}$ $(\Delta,pop_{50})$ & $0.610_{\pm0.298}$ & $0.588_{\pm0.241}$ & $5.212_{\pm1.987}$ \\
&$(\Delta,all)$ $\emph{v.s.}$ $(\Delta,pop_{100})$ & $0.664_{\pm0.299}$ & $0.667_{\pm0.262}$ & $4.314_{\pm1.829}$ \\
&$(\Delta,all)$ $\emph{v.s.}$ $(\Delta,uni_{10})$ & $0.772_{\pm0.206}$ & $0.709_{\pm0.252}$ & $2.909_{\pm1.347}$ \\
&$(\Delta,all)$ $\emph{v.s.}$ $(\Delta,uni_{50})$ & $0.830_{\pm0.166}$ & $0.758_{\pm0.212}$ & $2.124_{\pm1.199}$ \\
&$(\Delta,all)$ $\emph{v.s.}$ $(\Delta,uni_{100})$ & $0.838_{\pm0.164}$ & $0.800_{\pm0.214}$ & $1.768_{\pm1.088}$ \\
\bottomrule
\end{tabular}}
\end{table}

\paratitle{Results}. Table~\ref{t:analysis_sampling} presents the  correlation results 
of different sampled metrics, which are much smaller than those in Table~\ref{t:data_segmentation_result}.
It indicates that using sampled metrics has a larger influence on  performance ranking.
Indeed, such a concern has been discussed in recent studies~\cite{Rendle2020}: 
 sampled metrics are likely to be inconsistent and do not even persist qualitative ordering.
Another observation is that sampling more irrelevant items increases the correlation degree between the  sampled and exact metrics. 
 Finally, different sampling strategies are likely to cause the performance turbulence of some specific algorithms, which substantially affects the performance ranking. 
Comparing two sampling strategies,  it seems that uniform sampling is more closely correlated with the entire ranking. 
Generally, sampled metrics should not be used for small datasets. 
If  needed, we suggest using more irrelative items (\eg 1000 items suggested by \cite{Koren2008}). 

\subsection{Analysis on Domain Selection}

Above, we compute the correlation results by averaging over 23 different domains. Here, we consider whether   different domains lead to varying performance rankings. 
Such an issue is useful to answer how to select suitable  datasets for evaluation.  

\paratitle{Setting}. Given two domains, we first generate a configuration according to the suggested setting, \ie $(RO, RS, all)$, from 
Section 4.1 and 4.2,  and then obtain a performance ranking for each domain under the configuration for some metric. 
Then, we compute the \textit{Spearman's Rank Correlation} score  between  two domain-specific rankings.
We   average the \textit{SRC} scores over all the metrics. The final score is used to measure the  correlation between two domains.

\begin{figure}[!t]
\centering
\includegraphics[width=0.8\columnwidth]{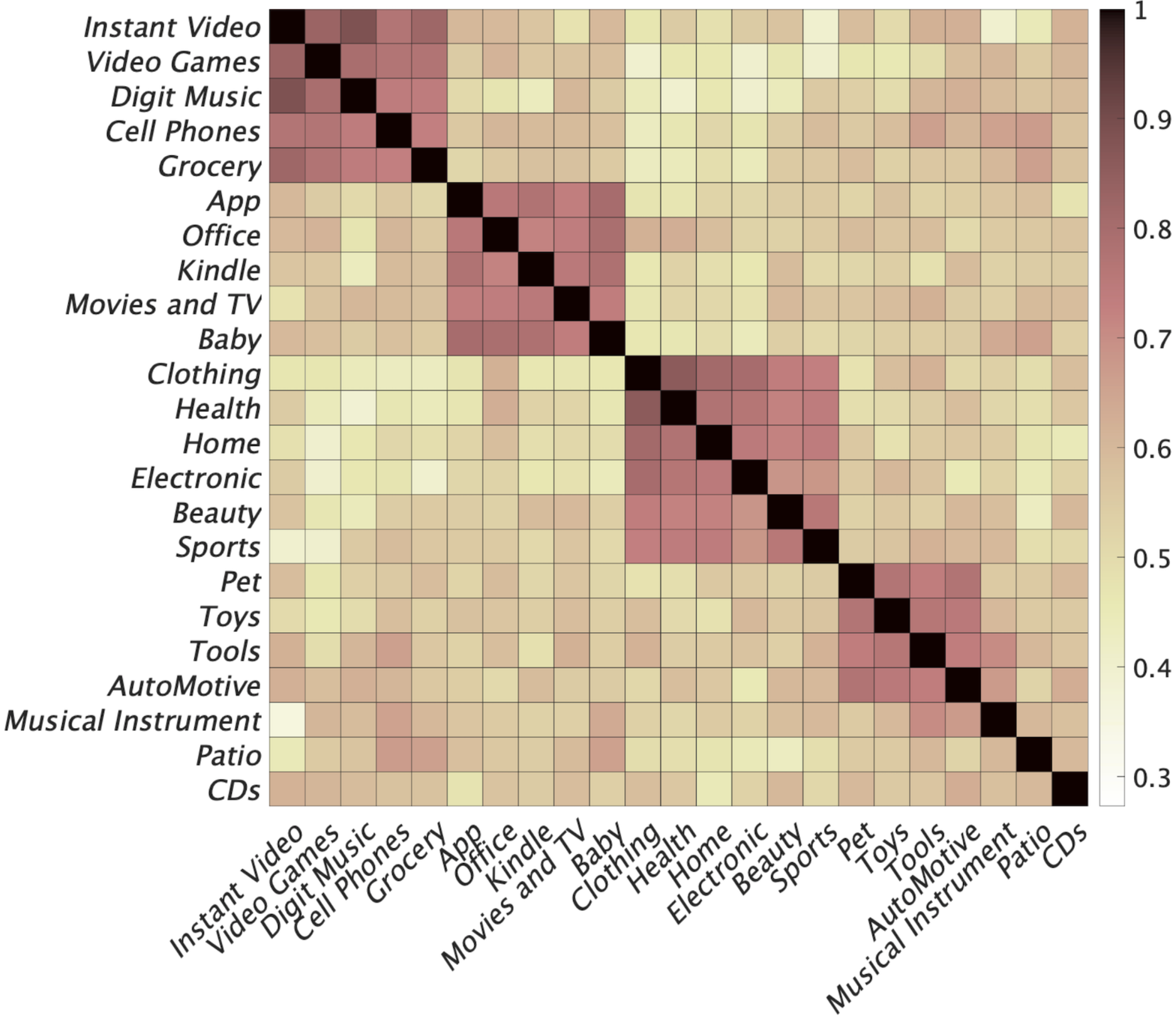}
\caption{Visualization of pairwise domain correlations. Each cell indicates the computed correlation score between two domains  (a darker color indicates a larger value). }\label{fig:case}
\end{figure}

\paratitle{Results}.  Figure~\ref{fig:case} presents the average correlation results between two domains. We reorder the rows or columns so that large values can be aggregated in the diagonal line. 
Interestingly, the entire heatmap seems to contain four major groups, in which within-group correlation values are higher than those across groups.
The results indicate that different domains are likely to yield varying performance rankings under the same configuration. Hence,  domain difference should be considered in  evaluation. By inspecting into the dataset, we find that  domain characteristics (\eg the first group mostly corresponding to digital products) and sparsity levels (\eg the ratio of user-item interaction) seem to have
 significant effect on the correlation results. 
A good  strategy is to use several datasets of varying sparsity levels from diverse  domains.
Here, ``domains''  refer to the categories of the Amazon dataset.
Although these domains come from the same platform, the finding has reflected the concern to some extent when selecting datasets for evaluation. 
 We will examine this issue using more datasets in future work.  


